\documentclass[12pt,aps,prb,reprint, superscriptaddress, showpacs, showkeys]{revtex4-1}

\usepackage[]{graphicx}
\usepackage{epsfig}
\usepackage{amsmath}
\usepackage{amsfonts}
\usepackage{amssymb}
\usepackage{color}
\usepackage{subfigure}
\usepackage{float}
\usepackage{bm}

\renewcommand{\vec}[1]{{\ensuremath{\bm{\mathrm{#1}}}}}
\usepackage{dsfont}
\newcommand{\mat}[1]{{\ensuremath{\mathds{#1}}}}

\newcommand{\mb}[1]{{\ensuremath{\mathbf{#1}}}}

\newcommand{\smJij}{J_{ij}}

\newcommand{\smJijT}{\mat{J}_{ij}}

\newcommand{\sms}{\mathbf{S}}

\usepackage{amsthm}
\newcommand{\etal}{\textit{et al.} }

\begin{document}
\title{The Landau-Lifshitz equation in atomistic models}
\author{M.O.A Ellis}
 \affiliation{Department of Physics, University of York,
York, YO10 5DD, United Kingdom}
\author{R.F.L. Evans}
 \affiliation{Department of Physics, University of York,
York, YO10 5DD, United Kingdom}
\author{T.A. Ostler}
 \affiliation{Department of Physics, University of York,
York, YO10 5DD, United Kingdom}
\author{J. Barker}
\affiliation{Institute for Materials Research, Tohoku University, Sendai, Miyagi 980-8577,
Japan}
\author{U.\ Atxitia}
\affiliation{Universit{\"a}t Konstanz, D-78457 Konstanz, Germany}
\author{O.\ Chubykalo-Fesenko}
\affiliation{Instituto de Ciencia de Materiales de Madrid, CSIC,
Cantoblanco, 28049 Madrid, Spain}
\author{R.W. Chantrell}
 \affiliation{Department of Physics, University of York,
York, YO10 5DD, United Kingdom}
%%\cortext[cor1]{Corresponding author. Tel.:+44(0)7766 989885, Fax:+44(0)1904 432284 }

%%---------------------------Start abstract--------------------------------------------------
\date{\today}

\begin{abstract}
The Landau-Lifshitz (LL) equation, originally proposed at the macrospin level, is increasingly used in Atomistic Spin Dynamic (ASD) models. The models are based on a spin Hamiltonian featuring atomic spins of fixed length, with the exchange introduced using the Heisenberg formalism. ASD models are proving a powerful approach to the fundamental understanding of ultrafast magnetisation dynamics, including the prediction of the thermally induced magnetisation switching phenomenon in which the magnetisation is reversed using an ultrafast laser pulse in the absence of an externally applied field. The paper outlines the ASD model approach and considers the role and limitations of the LL equation in this context. 
\end{abstract}

\maketitle
\section{Introduction}
Atomistic spin models have a long history, going back to the pioneering work of Binder~\cite{Binder} and co-workers in the 1970s. Typically these studies concentrated on the static properties of spin systems, particularly using Monte-Carlo methods to investigate the order/disorder phase transition and finite size effects in magnetic nanoparticles. The atomistic approach proved a powerful tool in the study of purely thermodynamic aspects of magnetic spin systems. 

However, the use of atomistic models for simulating magnetisation dynamics was, until recently,  rather limited and generally based on Monte-Carlo simulations of escape over energy barriers. Such an approach is hampered by the fact that the timesteps are generally not quantified. Time quantification was attempted by Nowak et al.~\cite{Uli_TQMC}, but this is successful only in the strong damping regime where the precession can be neglected. The study of dynamic phenomena
however was intrinsically limited  until the development of dynamic\cite{Kodama,Mitsumata} and stochastic
atomistic spin models.\cite{Uli1,Eric,Skubic}

In general the model of choice for magnetisation reversal studies is micromagnetics. The history of micromagnetics starts with a 1935
paper of Landau and Lifshitz on the structure of a wall
between two antiparallel domains, and several papers
by Brown around 1940. A detailed treatment of
micromagnetism is given by Brown in his 1963 book~\cite{Brown}.
For many years micromagnetics was limited to the
use of standard energy minimization approaches to
determine domain structures and classical nucleation
theory to investigate magnetization reversal mechanisms
in systems with ideal geometry. Arguably, the
current interest in micromagnetics arises from the
availability, from about the mid-1980s onward, of
large-scale computing power which enabled the study
of more realistic problems which were more amenable
to comparison with experimental data.
One important realization during this period
was the fact that, although micromagnetics can predict the nucleation fields for the magnetic system, due to the coexistence of different energy minima, multiple magnetization reversal paths are possible. 
Thus micromagnetics does not predict necessarily correctly the
state of the system after magnetization reversal.
Consequently, a lot of work has gone into the
development of dynamic approaches which use simulations
based on the Landau-Lifshitz equation of
motion. This is probably the technique in most
common use today. 

Dynamic calculations using micromagnetics have become ubiquitous, finding applications in fundamental investigations of reversal dynamics of magnetic materials. In addition micromagnetic models are vital to many industries, including the development of new generations of magnetic recording heads and media and permanent magnets. However, limitations of the micromagnetic approach are becoming increasingly apparent. Firstly, as magnetic materials become increasingly structured at the nanoscale to investigate new physical phenomena and create new functionality, the continuum nature of micromagnetics reaches the limits of validity. Secondly, although thermal activation can be introduced into the micromagnetic formalism, its applicability is strictly limited to low temperatures;  it is known that micromagnetic models greatly over-estimate the Curie temperature.\cite{Koch,Martinez}
This is a serious limitation in the investigation of ultrafast magnetisation dynamics, where temperatures up to and beyond $T_c$ can be achieved on the picosecond timescale. The natural evolution is toward dynamical approaches with atomistic resolution; such models are gaining increasing traction in dealing with the physics of ultrafast magnetisation processes and practical problems such as providing an understanding of Heat Assisted Magnetic Recording (HAMR).\cite{Terry,Rottmayer} Remarkably, the key to Atomistic Spin Dynamic (ASD) models is the use of the LL equation at the atomic level. Here we outline the basis of ASD models, review some recent simulations of ultrafast spin dynamics and consider the physical justification for the use of the LL equation at the atomistic level.

\section{Atomistic Spin Dynamic models}

The physical basis of the atomistic spin model is the localization
of unpaired electrons to atomic sites, leading to an
effective local atomistic magnetic moment, which is treated as a classical spin of fixed length. Ab initio calculations of the electron density~\cite{Schwarz}
show that in reality, even in ‘itinerant’ ferromagnets, the spin
polarization is well localized
to the atomic sites. Essentially
this suggests that the bonding electrons are unpolarised, and
after taking into account the bonding charge the remaining
d electrons
form a well-defined
effective localized moment on
the atomic sites. Nonetheless the assumption of classical spins leads to a fundamental discrepancy with experiments which will be discussed later.

The basis of ASD models, reviewed by Evans et al.~\cite{Richard} is a classical spin Hamilonian based on the Heisenberg exchange formalism. The spin Hamiltonian $\mathcal{H}$ typically has the form:
\begin{equation}\label{eq:spinhamiltonian}
    \mathcal{H} = \mathcal{H}_{\mathrm{exc}} + \mathcal{H}_{\mathrm{ani}} + \mathcal{H}_{\mathrm{app}},
\end{equation}  
with the terms on the RHS representing respectively the exchange, anisotropy and Zeeman terms. The exchange term is usually isotropic in spin space and the anisotropy term includes energies which are angular dependent. These can arise from crystalline anisotropies or magnetostriction and strains. The exchange term can also be anisotropic in some situations as will be discussed later.

The exchange energy for a system of interacting atomic moments is given by the expression
\begin{equation}
    \mathcal{H}_{\mathrm{exc}} = -\sum_{i\ne j} \smJij \sms_i \cdot \sms_j
\label{eq:HJij}
\end{equation}
where $\smJij$ is the exchange interaction between atomic sites $i$ and
$j$, $\sms_i$ is a unit vector denoting the local spin moment direction and $\sms_j$ is the spin moment direction of neighboring atoms. The unit vectors are taken from the actual atomic moment $\boldsymbol{\mu}_{\mathrm{s}}$ and given by $\sms_i = \boldsymbol{\mu}_{\mathrm{s}}/|\boldsymbol{\mu}_{\mathrm{s}}|$.  Due to the strong distance dependence of the exchange interaction the sum in Eq.~(\ref{eq:HJij}) is often truncated to include nearest neighbors only. This significantly reduces the computational effort while being a good approximation for many materials of interest. In reality the exchange interaction can extend to several atomic spacings~\cite{Mryasov:2005kj,Szunyogh:2011db}, representing hundreds of pairwise interactions.

In the simplest case the exchange interaction $\smJij$ is isotropic, meaning that the exchange energy of two spins depends only on their relative orientation. In more complex materials, the exchange interaction forms a tensor with components:
\begin{equation}
\smJijT = \begin{bmatrix} J_{xx}&J_{xy}&J_{xz}\\ J_{yx}&J_{yy}&J_{yz} \\ J_{zx}&J_{zy}&J_{zz} \end{bmatrix}\mathrm{,}
\label{eq:HJijM}
\end{equation}
which is capable of describing \textit{anisotropic} exchange interactions, such as two-ion anisotropy\cite{Mryasov:2005kj} and the Dzyaloshinskii-Moriya interaction (off-diagonal components of the exchange tensor). In the case of tensorial exchange, the exchange energy is given by the product:
\begin{equation}
    \mathcal{H}_{\mathrm{exc}} = -\sum_{i\ne j} \sms_i^{T}  \smJijT \sms_j\mathrm{.}
\label{eq:HJijSMS}
\end{equation}

We now proceed to consider two important factors in the use of ASD, firstly the process of determining, from first principles, the parameters of the spin Hamiltonian and secondly the introduction of spin dynamics, and the implications of the use of the LL equation at the atomistic level.

\subsection{Ab-initio calculation of spin Hamiltonian parameters: Multiscale approaches}
The  material parameters central to the spin Hamiltonian are the exchange interactions, anisotropy energies, and the magnitude of the spin moment. These can generally be found using two routes: (i) experimental measurements, either in a mean-field sense from macroscopic quantities such as the Curie point, or microscopically using neutron scattering, (ii) with a multiscale approach using
ab initio Density Functional Theory (DFT) calculations to parameterise the spin model. The ab initio approach is often preferable as it removes extrinsic factors from the parameters such as non-uniformity of an experimental sample and also provides a resolution (for example many exchange neighbours) which is hard to obtain experimentally even with Neutron scattering. Difficulties can arises such as in magnetic materials involving Rare-Earths, where the treatment of the $4f$ electrons is problematic in DFT. In this case an experimental parameterization becomes the most practical route, as was done by Ostler et al.~\cite{TomPRB} for amorphous GdFeCo alloys and Evans et al.~\cite{RichardNdFeB} for Nd$_2$Fe$_{14}$B alloys.

Contemporary ab initio methods enable the calculation of a wide range of material properties including ground state magnetic properties. So-called `beyond DFT' methods allow the calculation of even small energy differences, providing access to the magnetic crystalline anisotropy constants. Standard software packages such as VASP\cite{Kresse} and SIESTA~\cite{Soler} make such calculations accessible to interested researchers. 
The calculation of pair wise exchange interactions in DFT is somewhat complicated by the delocalised nature of the basis sets employed. Parameterizing a spin Hamiltonian therefore requires mapping many different spin configurations onto the atomic Hamiltonian. An alternative is to use scattering methods such as Korringa-Kohn-Rostoker~\cite{Zabloudil,Ebert} or linear muffin tin orbitals~\cite{Misha,Pajda}. These methods are built around the atomic sphere approximation which gives a natural mapping onto the localised Heisenberg formalism in conjunction with the
magnetic force theorem~\cite{Liechtenstein}.

Connecting the ab-initio and atomistic length scale is an important link in the multiscale modelling chain allowing one to include both dynamics and temperature, which has been demonstrated for FePt by Kazantseva et al.~\cite{Natalia}. In this section we  consider the calculation of the temperature dependence of static materials properties. Mryasov et al.~\cite{Oleg} carried out ab-initio calculations of exchange and 
anisotropy of the L$_1$0 phase of FePt; an important candidate for HAMR media. The aim was to investigate the exchange and anisotropy values of bulk FePt and to map them onto a classical spin model in order to investigate  static and dynamic properties. This process was complicated by the induced Pt moments arising from the Fe exchange field. Such an effect is beyond the Heisenberg moment of fixed moments, but based on ab-initio calculations an effective spin Hamiltonian was developed which is dependent only on the Fe degrees of freedom. 

\begin{equation}
  \mathcal{H}^{\mathrm{FePt}} = - \sum_{i \ne j} \big( \tilde{J}_{ij} \vec{S}_i \cdot \vec{S}_j +
        \tilde{D}^{(2)}_{ij} S^z_i S^z_j\big)  - \sum_i \tilde{D}^{(0)}_i (S^z_i)^2\mathrm{.}
\label{eq:newHam1}
\end{equation}
The exchange parameters include the effect of Fe-Pt-Fe interactions which contributes to both the isotropic exchange
$\tilde{J}_{ij}$ as well as introducing a two-ion anisotropy, $\tilde{D}^{(2)}_{ij}$, because of the layered ordering of this intermetallic. This two-ion term is considerably larger than the single ion anisotropy $\tilde{D}^{(0)}_i$. The exchange interactions $\tilde{J}_{ij}$ are significant over large distances, making numerical calculations rather time consuming. Physically it leads to strong finite size effects\cite{Ellis2015} and in particular leads to deviations of the finite size scaling exponents from the expectations of the nearest neighbour Heisenberg form~\cite{Ondrej,Andreas1,Andreas2}.

In ref.~\onlinecite{Oleg}, the thermodynamic properties of FePt were investigated using an atomistic model based on the spin Hamiltonian given in Eq. (\ref{eq:newHam1}). It was shown that the two-ion term gives rise to a thermal anisotropy scaling exponent $n=2$ ( with $K(T) \propto M^n(T)$)
%(with $K(T)/K(0)=(M(T)/M(0))^n$) 
consistent with the theory of Callen and Callen~\cite{Callen} and in contrast to the single ion anisotropy for which the exponent is $n=3$.  The importance of the atomistic approach is that it is able to calculate the exact exponent arising from the specific material parameters of FePt. This resulted in $n=2.1$, the non-integer value reflecting the relative importance of the single- and two- site anisotropy terms; a value in good agreement with experiments on FePt nanoparticles~\cite{Thiele,Okamoto}.

\subsection{Langevin Dynamics and the LL equation at the atomic level}
The important step forward in the use of atomistic models is the introduction of Langevin Dynamics, allowing modelling of the dynamical response of the magnetisation to temperature changes~\cite{LD}. The approach is based on the introduction of thermal fluctuations for a single particle developed by Brown~\cite{Brown}. The theoretical basis is the classical
theory of Brownian motion which accounts for
the departure from thermal equilibrium due to the energy
interchange between a particle and its heat bath, with the
Landau-Lifshitz equation augmented by white-noise fields, effectively producing the Stochastic (Langevin) equation of the problem. The approach is to determine local fluctuating fields using the fluctuation dissipation
theorem and to require the equilibrium distribution
function of the orientations of the magnetization to
coincide with the Boltzmann distribution.

In order to yield the Boltzmann equilibrium distribution, the stochastic LL equation should be interpreted as a Stratonovich vector stochastic differential equation~\cite{Garcia}. This is integrated by a suitable choice of the numerical integration scheme, most usually that of Heun. Care must be taken that the spin moments remain of unit length and for a non-conservative scheme such as Heun, an explicit renormalization of the length at each timestep is required for the Stratonovich solution~\cite{Berkov}. The integration of the stochastic LL equation is discussed in detail in ref.~\onlinecite{Richard}

Consequently, the basis of ASD for a set of coupled spins
is the integration of the stochastic Landau-Lifshitz equation for each
localized magnetic moment $\vec{S}_i$:
 \begin{equation}
 \label{LLG}
\mb{\dot{S}}_i=\gamma[\mb{S}_i\times\mb{H}_i]-\gamma\alpha[\mb{S}_i\times[\mb{S}
_i\times\mb{H}_i]]
\end{equation}
Here $\mb{H}_i = \vec{\xi}_i(t) -\partial \mathcal{H}/\partial\vec{S}_{i}$ is the local effective field which includes Zeeman,
exchange, anisotropy and magnetostatic contribution, augmented with
by a stochastic term $\vec{\xi}_i(t)$ (which appears like a field). It is defined through the correlators:
\begin{equation}
\label{whitenoise}
     \left\langle \xi_i(t) \right\rangle=0, \ \  \
\left\langle\xi_{i\eta}(t)\xi_{j\nu}(t')\right\rangle=
     \frac{2\alpha k_B T}{\gamma \mu_{s}}\delta(t-t')\delta_{ij}\delta_{\eta\nu}.
\end{equation}
Here $T$ is the temperature of the heat bath, $\gamma$ is the gyromagnetic ratio, $\mu_{s}$ is the magnetic moment, $\alpha$ is the  parameter describing the coupling strength to the heat bath, $\eta$ and $\nu$ are Cartesian components. The basis of this equation is the separation of timescales, assuming that
the bath (phonon or electron system) is much faster than the spin system. Consequently, the fluctuation-dissipation theorem can be applied to derive the equilibrium
white noise properties of Eq.~(\ref{whitenoise}). In section~\ref{analysis} we consider the applicability of the Langevin Dynamic approach using the LL equation. However, prior to this we give some examples of the success of ASD models in developing an understanding of the thermodynamic aspects of ultrafast magnetisation processes, including the prediction of a novel `linear' reversal mechanism.

\section{Atomistic models of ultrafast spin dynamics}
Advanced models are required in order to understand ultrafast magnetisation processes. Interest in this area has developed rapidly since the experimental demonstration~\cite{Beaurepaire} that the magnetisation of Ni can be reduced by laser heating on a timescale of 1ps. Experimentally the measurements are made using a pump-probe process.  A high energy femtosecond laser is used to heat the material (pump), and the magnetic response is measured using MOKE with a low energy probe beam split off from the pump. This experiment gives time resolved measurements of the magnetic response following pulsed laser heating. Such experiments are extremely challenging in terms of understanding the physics of ultrafast magnetisation processes and damping mechanisms. 

A more recent development was the experimental demonstration by Stanciu et al.~\cite{Stanciu}, of optically induced magnetisation reversal in the amorphous ferrimagnet GdFeCo. Using circularly polarised ultrafast laser pulses, Stanciu et al. showed magnetisation reversal to be dependent on the chirality of the laser pulse. 
This was interpreted as arising from a large, laser-generated, optomagnetic field
(estimated as large as 20T) possibly originating from the Inverse Faraday Effect. 
The reversal was explained~\cite{Khadir} using atomistic and Landau-Lifshitz-
Bloch (LLB) macrospin simulations (for a review of the LLB equation see~\cite{Oksana}) as arising from this large optomagnetic field together with purely thermodynamic contribution which initiates switching via the so-called linear reversal mechanism~\cite{Natalia_lin}.
Linear reversal is an important prediction of the atomistic model and property of the LLB equation. Essentially it involves a collapse of the magnetisation to zero and subsequently a switched polarisation in  a reversing field. Linear reversal sets in at a critical temperature $T^*$ related to the ratio of the longitudinal and transverse susceptibilities~\cite{Natalia_lin}. Importantly, this leads to ultrafast magnetisation reversal since the timescale is governed by the longitudinal relaxation time which is in the order of hundreds of femtoseconds.

\begin{figure}[htb!]
\includegraphics[width=\columnwidth]{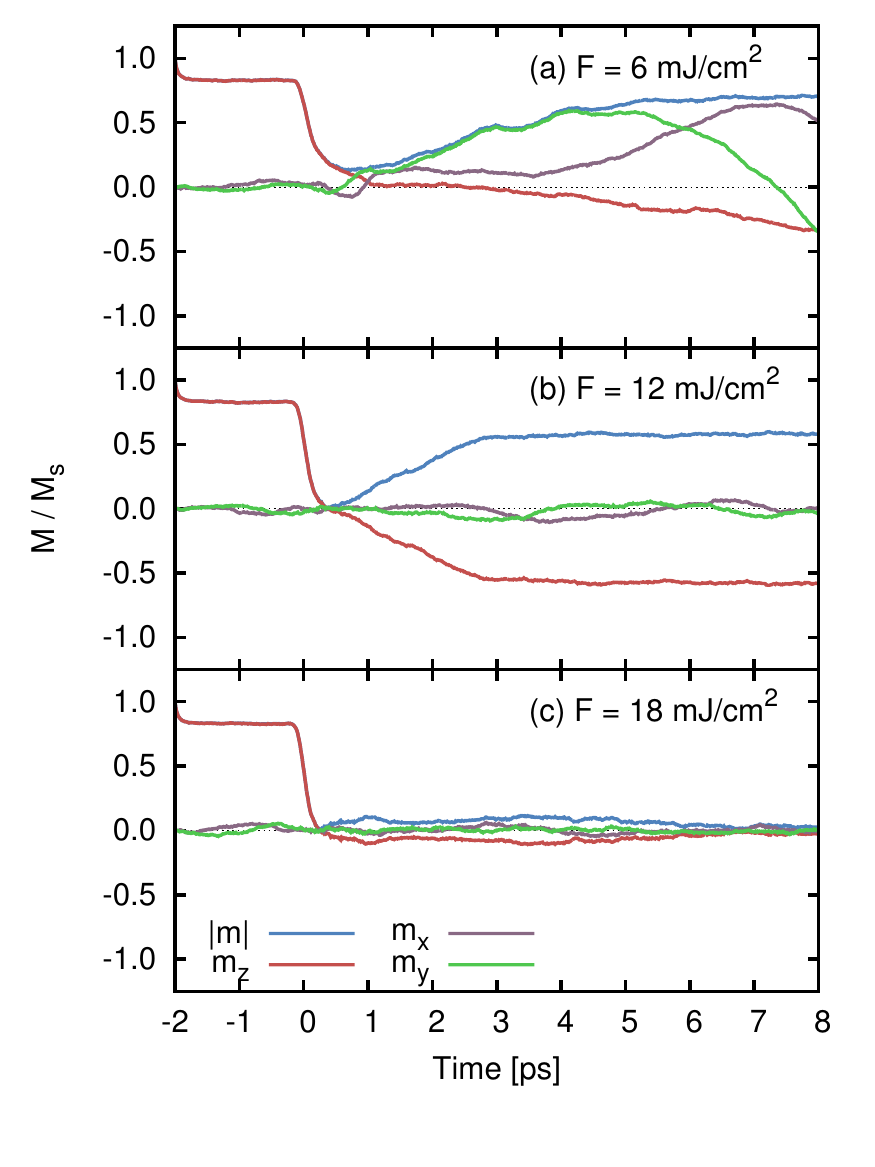}
\caption{The temporal evolution of the magnetization in FePt after a femtosecond laser pulse simulated using the atomistic model described in section II. The model uses the Hamiltonian expressed in Eq. \ref{eq:newHam1} with a damping, $\alpha = 0.1$, and $\mu_s = 3.23 \mu_B$. The system starts at room temperature before excitation by a 100 fs laser pulse incident at t=0 ps with fluences of (a) 6 mJ/cm$^2$ (b) 12 mJ/cm$^2$ and (c) 18 mJ/cm$^2$. A reversing optomagnetic field also occurs with the laser pulse; it has a square pulse shape from t=0 ps to t = 1 ps with a magnitude of 30 T.}
\label{fig:linear}
\end{figure}

Originally optomagnetic switching was only observed in various ferrimagnetic structures but it can, in principle, occur in ferromagnets as well. The experimental work by Lambert \etal\cite{Lambert2014a} found that thin films of ferromagnetic Co/Pt layers and granular FePt showed magnetisation switching after repeated excitation by a circularly polarised femtosecond laser. Fig. \ref{fig:linear} shows an atomistic model simulation of the response of the magnetisation of FePt to the field and temperature pulses associated with a femtosecond laser pulse. The calculations used the spin Hamiltonian of Mryasov et al.\cite{Oleg} given in Eq.~(\ref{eq:newHam1}) with a optomagnetic field of 30T that lasts for 1 ps after laser excitation. The dynamic response is calculated using Langevin Dynamics driven by the electron temperature evolved using a 2-temperature model~\cite{2T}. The reversal depends critically on the laser power, as shown in Fig.~\ref{fig:linear}. At low laser power, Fig. \ref{fig:linear}(a), demagnetisation is not complete and the reversal proceeds via the usual precessional route, albeit over an energy barrier reduced due to the quenching of the anisotropy. Complete reversal proceeds over many picoseconds. At elevated temperature, Fig. \ref{fig:linear}(b), switching of the total magnetisation proceeds by a process involving no macroscopic precession; the linear reversal mode. Importantly, switching via the linear reversal mode occurs on a timescale of the longitudinal relaxation time of the magnetisation ($\sim$hundreds of fs). At higher laser power, Fig. \ref{fig:linear}(c), reversal occurs but the magnetisation is destroyed by the elevated temperature. Vahaplar et al.\cite{Khadir} show that optically induced switching proceeds above a critical temperature which is sufficient to excite linear reversal but not so large as to demagnetise the system.

It was shown by Vahaplar et al.~\cite{Khadir} that extremely high optically-induced  fields (tens of Tesla) needed to be invoked to trigger the all-optical reversal. The possible origin of such fields remains a matter of debate. However, it can be shown that the fields may arise from interatomic exchange forces. This interpretation begins with the observation by Radu et al.~\cite{Radu} that the RE and TM sublattices demagnetise at different rates, even though strongly coupled through intersublattice exchange forces. The measurements were made using XMCD to provide the element specific magnetisation dynamics. The experimental observations were supported by atomistic model calculations, which verified both the differential sublattice dynamics and the existence of an intriguing transient ferromagnetic-like state (TFMLS) which is created by the reversal of the TM spins into the RE spin direction. The TFMLS exists for around 500fs and is associated with the magnetisation reversal process. Further investigation led to the astonishing prediction by Ostler et al.\cite{Tom}, using atomistic model calculations that switching occurred \emph{ in the absence of any external symmetry breaking field.} In ref.~\onlinecite{Tom} this remarkable prediction was verified experimentally. Interestingly, thermally induced magnetisation switching (TIMS) allows a re-interpretation of optically-induced magnetisation reversal. 
Rather than invoking large fields of opto-magnetic origin, Khorsand et. al.~\cite{Khorsand}  gave experimental evidence in favour of TIMS as the main switching mechanism in GdFeCo and attribute the helicity dependence of the laser excitation on the dichroic effect, i.e. the dependence of the absorption of energy from the laser pulse on the chirality of the laser light.

The phenomenon of TIMS has been further investigated theoretically using macrospin models~\cite{Mentinck,Unai} and atomistic approaches~\cite{Joe}. Importantly, refs.~\onlinecite{Unai,Joe} show that TIMS arises from angular momentum transfer between the RE and TM sublattices mediated by the establishment of a two-magnon bound state. The study in ref.~\onlinecite{Joe} involved detailed calculations of the magnon band structure, which has two branches with properties which strongly depend on the material composition. Calculations showed that transfer of energy between the modes, resulting from laser excitation, was the physical origin of the TIMS phenomenon. Further, the calculations demonstrated a window for TIMS within a certain range of alloy concentrations. Specifically, for low RE concentration essentially a uniform FM mode is excited. With increasing RE concentration, the optical mode becomes accessible, leading to TIMS. At higher RE concentrations there develops a large band gap which precludes the angular momentum transfer between sublattices, at which point TIMS cannot be excited. This prediction is in good agreement with experiment. 

The results of ref.~\onlinecite{Joe} allow the definition of design rules for TIMS. Specifically,
\begin{enumerate}
\item The existence of two sublattices with differential demagnetisation dynamics
\item Antiferromagnetic coupling between the sublattices
\item Reversal of the dominant magnetic sublattice which stabilises the switched magnetisation direction during the cooling phase following the laser pulse.
\end{enumerate}
This gives rise to the expectation that synthetic ferrimagnetic (SFIM) structures consisting of two ferromagnetic layers separated by an element such as Ru or Ir, to establish AF coupling between the layers, would exhibit TIMS. This has been demonstrated numerically by Evans et al.~\cite{SFIM} in bilayers of Fe and FePt coupled by an exchange separation layer assumed to promote AF coupling between the FM layers. Importantly, this extends the TIMS phenomenon beyond its initial prediction and discovery in amorphous ferrrimagnets~\cite{Tom} to designed materials with high anisotropy and the avoidance of RE components. This is of importance both in terms of fundamental understanding of ultrafast magnetisation processes and of applications in information storage. In the latter context we note that Evans et al.\cite{Limit} demonstrated the requirement for large write fields in the magnetic recording process; not simply to ensure magnetisation reversal, but also that there is no back-switching of the magnetisation, which would lead to a limiting source of noise. Ostler et al.~\cite{Tom} show that extremely large fields $\sim$ tens of Tesla are necessary to oppose the formation of the TFMLS. This is consistent with the physical origin of TIMS in the excitation of the two-magnon bound state, which naturally introduces fields arising from the exchange interaction.

Clearly, TIMS is an important prediction using a model based on LL dynamics at the atomistic level and shows the LL equation to be a remarkable piece of insight which finds application at time- and length- scales significantly beyond its original conception. In the following we consider the underlying physics of the LL equation at the atomistic level, distinguishing between the origin of the precession and damping terms.

\section{Analysis of the Atomistic Langevin Dynamic approach using the LL equation}
\label{analysis}
The precession term in the LL equation derives directly from the equation of motion
\begin{equation}
i\hbar \frac{d}{dt} \langle M \rangle = \langle[M,\hat{H}] \rangle,
\end{equation}
with $\hat{H}=-M.B(t)$, which leads directly to the larmor precession term.
The damping term is more difficult to justify at the atomistic level. Pragmatically one can make the case that some term coupling the spin to the heat bath is necessary to ensure eventual thermal equilibrium with the heat bath. In this spirit, the LL damping term is the simplest form capable of including this key physical requirement. Some work has been carried out to investigate spin dynamics at the quantum level. In particular, Cappelle and Gyorffy~\cite{Capelle} have investigated magnetisation dynamics using time dependent density functional theory. They construct a gradient-dependent density functional which is then used to derive the
phenomenological LL form of damping directly from first principles. 
The LL equation and its damping term can also be derived from the quantum mechanics using the density matrix formalism~\cite{Blum}, an approach recently re-visited by Weiser~\cite{Wieser}.
The coupling of the spin to the heat bath is expected to be material dependent; essentially it represents the relevant energy transfer channel which might arise from interaction between the spin and conduction electron or spin/lattice interactions and should be studied at the quantum level. 

The introduction of thermal fluctuations into the atomistic model formalism is achieved via the Langevin Dynamic approach. As mentioned previously, the assumption is made that the instantaneous random fields, by which the thermal fluctuations are introduced, are uncorrelated in time and space, giving rise to white noise.
The basis of this formalism is the separation of timescales, assuming that the bath (phonon or electron system) is much faster than the spin system. In this case, the bath degrees of freedom can be averaged out and replaced by a stochastic field with white noise correlation functions.
In sense this is perhaps the most questionable part of the application of the atomistic model in ultrafast magnetism, which involves phenomena on the timescale of tens of femtoseconds, where spatial and temporal noise correlations might be expected. 

Investigation of this phenomenon is the province of models simulating the thermal behaviour of the heat bath and coupling this to the spin system. In this case only the precession term of the LL equation is retained, with the damping included via a term based on the underlying mechanism coupling the spin to the heat bath. This was treated in a generic way by Atxitia et al.~\cite{Unai_noise} using an approach developed by Miyazaki and Seki\cite{MandS} and generalising this to multi-spin systems. The noise takes the form of an Ornstein-Uhlenbeck process~\cite{Sancho} introducing a characteristic correlation time $\tau_c$.
The spin is coupled locally to the bath which is connected to a thermostat as follows
\begin{eqnarray}
\label{MS}
\mathbf{\dot{S}}_i&=&\gamma [\mathbf{S}_i \times (\mathbf{H}_i + \boldsymbol{\eta}_i)], \nonumber \\
\boldsymbol{\dot{\eta}}_i&=&-{1 \over \tau_c}(\boldsymbol{\eta}_i - \chi \mathbf{S}_i)+ \mathbf{R}_i
\end{eqnarray}
with the fluctuation-dissipation theorem for the bath variable:
$\left\langle\mathbf{R}_i(t)\right\rangle=0$ ; $\left\langle\mathbf{R}_i(t)\mathbf{R}_j(t')\right\rangle=(2\chi k_B T /
\tau_c)\delta_{ij}\delta(t-t')$. The parameter $\chi$ describes
the coupling of the bath variable to the spin. The precession term in
the first equation of the set (\ref{MS})  has the same form as in
the Eq.(\ref{LLG}). However, the damping is now described by the
second equation in this set. In the second equation in Eqs.~(\ref{MS}) the bath variable
adjusts to the direction of the spin due to the interaction with it.
In the limit $\tau_c
\rightarrow 0$  the stochastic LL equation (\ref{LLG}) is
recovered. This also provides a relation between the damping  and
the coupling constants as $\alpha=\gamma \chi \tau_c $, giving
a more precise physical sense to the LL damping constant at the
atomistic level. 

In this approach, the phenomenological LL damping parameter is substituted by two unknown parameters: the correlation time $\tau_c  $ and the coupling constant $\chi$. Several processes may be
important in determining these constants, for example,
the spin-orbit coupling, momentum relaxation, scattering
rate and dephasing time of conduction electrons. As in the
LL approach, these parameters will be material specific
and their physical origins should be clarified on the basis of
first-principle approaches.
The effect of the correlation time on the ultrafast demagnetisation process was investigated in ref.~\onlinecite{Unai_noise}.  
For correlation time $\tau_c < 1$ fs  the correlated approach gives the same results as the  standard Langevin Dynamics with white noise 
However, in the range $\tau_c \simeq 10-100$fs the correlations were found to give a dramatic increase of the
longitudinal relaxation time. The effect is less pronounced at
higher temperature since in this case the temperature contributes to
the loss of correlations. Calculations based on the Langevin Dynamic approach generally give reasonable values for the longitudinal relaxation time in comparison with experiment, which suggests that experimental correlation times are on the order of 10 fs or less; greater values would have an appreciable effect on the observed rates of demagnetisation.

The interactions of the spins with the lattice system also provides
 an energy channel for the fluctuation and dissipation. In conventional
atomistic models the lattice is fixed and so transfer to and from the lattice is
handled phenomenologically by the fluctuation and damping terms in the stochastic
LL equation. By introducing the motion of the atoms to the model the energy transfer 
can be modelled directly without the need for the phenomenological terms. Recently 
models such as this have been developed to investigate a variety of systems where
the spin-lattice effects are important. 
Ma \etal\cite{Ma2008} have extensively developed a spin-lattice model of Fe that utilises
the dependence of the exchange on the atomic separation as the coupling between
the spins and lattice. However in this case both the spins and lattice use a
Langevin thermostat to maintain a constant temperature. Through this model both
systems act as a thermal reservoir from which  instantaneous spin and lattice temperatures 
can be extracted which can then be dynamically linked to the
electron temperature thus representing a dynamic three temperature model.\cite{Ma2012a}
Using this model, termed Spin-Lattice-Electron Dynamics (SLED) by Ma \etal, experimental
ultrafast laser induced magnetisation dynamics can be 
 reproduced.

Whilst the exchange interaction can couple the spins and lattice further coupling terms
have also been investigated. Karakurt \etal\cite{Karakurt2007} implemented a spin-lattice model
where the exchange is constant but introduce a specific coupling term of the form:

\begin{equation}
	\mathcal{H}_{\text{c}} = - C \mathbf{S}_i \cdot \mathbf{r}_{ij}
\end{equation}
 Where $\mathbf{r}_{ij} = \mathbf{r}_i - \mathbf{r}_i$ is the separation of 
the atoms and $C$ is a parameter to control the strength of the coupling.
Using this Karakurt \etal  demonstrated that this coupling causes a damping of 
the uniform precession mode.
Beaujouan \etal\cite{Beaujouan2012} propose a different type of coupling based on a two site anisotropy 
which takes the form of pseudo-dipole interaction. 
\begin{equation}
	\mathcal{H}_{\text{c}} = - K(\mathbf{r}_{ij}) \left( \left(\mathbf{\hat{r}}_{ij} \cdot \mathbf{S}_i \right) \left(\mathbf{r}_{ij} \cdot \mathbf{S}_j \right) - \frac{ \mathbf{S}_i \cdot \mathbf{S}_j}{3} \right)
\end{equation}
In this case the coupling strength, K, depends on the separation of the atoms and requires 
specific parametrisation from \textit{ab initio}. As discussed in the literature this form
of coupling arises from the spin-orbit interaction 
of the electrons and thus is more physically justifiable but is still not exact.
With this Beaujouan \etal are able to show that energy can be transferred between
the systems and an equilibrium temperature is obtained.
It is clear that by incorporating lattice dynamics into the ASD model various
effects which are treated phenomenologically are present. However these
still require a high level of empirical parametrisation for both the atomic
bonding and the spin-lattice coupling.

Finally, we consider briefly one further aspect of the use of fixed spin models such as the LL equation, specifically the classical nature of the spin. This leads to a disparity between the simulated and  experimental temperature dependent magnetization curves~\cite{KuzminPRL2005}. At the macroscopic level the temperature dependent magnetization is well fitted by the phenomenological equation proposed by Kuz'min\cite{KuzminPRL2005}. However, the Kuz'min equation merely describes the form of the curve with little relation to the microscopic interactions within the material which determine fundamental properties such as the Curie temperature. Ideally one would perform 3D quantum Monte Carlo simulations\cite{SandvikPRB1991}, but, although this is possible for small numbers of atoms, for larger ensembles the multiscale approach using atomistic models parameterized with ab-initio information remains the only feasible approach to connect the quantum and thermodynamic worlds. 
Evans et al.\cite{Richard_scaling} have proposed a scaling approach which maps classical to quantum spin models. The scaling recognises that, although the classical treatment finds the correct magnon energies, the distinction between classical
and quantum models results from the particular statistical properties of each approach. While quantum systems obey Bose-Einstein statistics, leading to the Bloch $T^{3/2}$ law at low temperatures, the classical Boltzmann statistics gives rise to a finite slope of the magnetisation at low temperatures. 
In ref.~\onlinecite{Richard_scaling} the existence of a simple relation between classical and quantum temperature-dependent magnetization at low temperatures is demonstrated. The temperature-dependent magnetization is represented in the
simplest form arising from a straightforward interpolation of
the Bloch law and critical behaviour given by the
Curie-Bloch equation
\begin{equation}
m(\tau) = \left(1 - \tau^{\alpha} \right)^{\beta}
\label{eq:mvsT}
\end{equation}
where $\alpha$ is an empirical constant and $\beta \approx 1/3$ is the critical exponent. Evans et al.\cite{Richard_scaling} then use the classical spin model simulations to determine the critical exponent $\beta$ and then find $\alpha$ by fitting the classical model predictions to experimental data. This leads to a mapping from a `simulation temperature' to the real temperature. It was shown that this approach gives excellent agreement with experiment\cite{Beaurepaire} for the demagnetisation of Ni following an ultrafast laser pulse.
\section{Conclusion}
The use of the LL equation in ASD models of magnetic materials has been described. The introduction of the LL equation, in its stochastic form, is the basis of a powerful approach to ultrafast sin dynamics. In particular, ASD simulations demonstrate the important thermodynamic contribution to laser-induced ultrafast processes. The models demonstrate a new, so-called linear reversal mechanism which is the path to ultrafast reversal. Also predicted is the phenomenon of TIMS, which is currently under extensive investigation and holds the promise of application in future devices requiring fast switching. The success of the LL equation in this framework is remarkable. While further investigations of the energy transfer mechanisms at the quantum level should be carried out to improve the physical understanding of damping, ASD methods based on the LL equations are likely to have an important role in understanding the physics of magnetic phenomena, not only at short timescales and elevated temperatures, but also on lengthscales where the micromagnetic formalism is not appropriate.

\section*{Acknowledgement}
Financial support of the EU Seventh Framework Programme under grant
agreement No.~281043, FEMTOSPIN is gratefully acknowledged.

\end{document}